\documentclass[a4paper,10pt,twoside]{cpc-hepnp}

\usepackage{multicol}
\usepackage{graphicx}
\usepackage{booktabs}
\usepackage{amssymb,bm,mathrsfs,bbm,amscd}
\usepackage[tbtags]{amsmath}
\usepackage{lastpage}

\begin{document}

\fancyhead[c]{\small Chinese Physics C~~~Vol. xx, No. x (201x) xxxxxx}
\fancyfoot[C]{\small 010201-\thepage}


\title{Mass spectra and Regge trajectories of $\Lambda_{c}^{+}$, $\Sigma_{c}^{0}$, $\Xi_{c}^{0}$ and $\Omega_{c}^{0}$ Baryons}

\author{%
      Zalak Shah$^{1}$
\quad Kaushal Thakkar$^{2}$
\quad Ajay Kumar Rai$^{1;1)}$\email{raiajayk@gmail.com}%
\quad P. C. Vinodkumar$^{3}$
}
\maketitle

\address{%
$^1$ Department of Applied Physics, Sardar Vallabhbhai National Institute of Technology, Surat, Gujarat, India-395007\\
$^2$ Department of Applied Sciences \& Humanities, GIDC Degree Engineering college, Abrama, Navsari, India-396406\\
$^3$ Department of physics, Sardar Patel University, V.V. Nagar, Anand, India-388120 \\
}

\begin{abstract}
We calculate the mass spectra of the singly charmed baryons ($\Lambda_{c}^{+}$, $\Sigma_{c}^{0}$, $\Xi_{c}^{0}$ and $\Omega_{c}^{0}$) using Hypercentral constituent quark model(hCQM). The hyper color coloumb plus linear potential is used to calculate the masses of positive(upto $J^{p}=\frac{7}{2}^{+}$) and negative parity(upto $J^{p}=\frac{9}{2}^{-}$) excited states. The spin-spin, spin-orbital and tensor interaction terms are also incorporated for mass spectra. We have compared our results with other theoretical predictions and Lattice QCD for each baryons. Moreover, the known experimental results are also reasonably closed to our predicted masses.  By using the radial and orbital excitation, we construct Regge trajectories for the baryons in (n,$M^{2}$) plane and find their slopes and intercepts. The other properties like, magnetic moments, radiative transitions and radiative decay widths of these baryons are also calculated successfully.  
\end{abstract}



\footnotetext[0]{\hspace*{-3mm}\raisebox{0.3ex}{$\scriptstyle\copyright$}2013
Chinese Physical Society and the Institute of High Energy Physics
of the Chinese Academy of Sciences and the Institute
of Modern Physics of the Chinese Academy of Sciences and IOP Publishing Ltd}%

\begin{multicols}{2}

\section{Introduction}
The measurement and calculation of baryonic excited states are the important area of activity for worldwide experimental facilities in CLEO, Belle, BABAR, LHCb \cite{CDF,cleo,2, 3,{belle}} as well as of Lattice QCD calculations \cite{29,4}. Baryons made of light and heavy flavors quarks put an interesting challenges because one has to deal with quark mass at various mass scale. Rapid development has been made in the  heavy baryons sector and many new excited states of these heavy flavour baryons are also investigated \cite{olive}. The doubly and triply heavy flavor baryons are determined now a days with different approaches \cite{v15,kei,t14,88}. In creating the understanding towards the different energy scale QCD, it is very much required a rich dynamical study on heavy flavor baryons with their properties \cite{galkin}. The mass spectra, width, lifetime, decays and form factors have been often reported by numerous experimental groups but the spin and parity identification of some states are still missing. The future experiments at J-PARC, PANDA \cite{panda} and LHCb are expected to give further information on charmed baryons pretty soon.\\

Various phenomenological models have been used to study heavy baryons by different approaches like non-relativistic Isgur-Karl model \cite{Isgur}, relativized potential quark model \cite{Capstick}, relativistic quark model \cite{ebert2011}, variational approach \cite{Roberts2008}, the relativistic flux tube (RFT) model \cite{chen2015}, the Hamiltonian model \cite{yoshida}, heavy quark symmetry \cite{yama}, the Fadeev approach \cite{valcarce}, the algebraic approach \cite{9}, the Goldstone Boson Exchange Model \cite{6}, the interacting quark-diquark model \cite{8}, the Feynman-Hellmann theorem \cite{28}, the Hypercentral Model  \cite{5}, the combined expansion in $1/m_Q$ and $1/N_c$ \cite{100}, the chiral Quark model \cite{99}, QCD sum rules \cite{98}, a soliton model \cite{90} and many more have been studied henceforth. Also, many Lattice QCD studies have been done for heavy flavor baryons \cite{4,brown,pacs,alex,mathur,29,lattice}. 
In all the models they first reproduced the baryon spectra and later described various baryonic properties. A recent Review article  explained different quark models along with the Hypercentral Constituent Quark Model and its application to baryonic properties \cite{ginnani2015}.\\

\begin{table*}
\caption{\label{tab:table1} The possible states for these experimentally known ($J^{P}$) states.}
\begin{tabular}{cc|cc|cc|cc}
\toprule
Names & Mass & Names & Mass &  Names & Mass &  Names & Mass \\ 
\hline 
$\Lambda_{c}(2286)^{+}$ &2286.46$\pm$.014  & $\Sigma_{c}(2455)^{0}$ & 2453.74$\pm$0.16 &$\Xi_{c}(2468)^{0}$ & 2470.88  & $\Omega_{c}(2695)^{0}$ & 2695.2 $\pm$ 1.7\\ 
 
$\Lambda_{c}(2595)^{+}$ &2592.25$\pm$.028 & $\Sigma_{c}(2520)^{0}$ & 2518.8$\pm$.06 & $\Xi_{c}(2645)^{0}$ &2645.9$\pm$0.5& $\Omega_{c}(2770)^{0}$ & 2765.9 $\pm$ 2.0\\ 

$\Lambda_{c}(2625)^{+}$ &2628.11$\pm$0.19 & $\Sigma_{c}(2800)^{0}$ & $2806 \pm8\pm10$ &$\Xi_{c}(2790)^{0}$  & 2791.8$\pm$3.3 & • \\ 

$\Lambda_{c}(2880)^{+}$ &2881.53$\pm$.035 & & & $\Xi_{c}(2815)^{0}$ & 2819.6$\pm$1.2  & • &  \\ 

$\Lambda_{c}(2940)^{+}$ &2939.3$^{+1.4}_{-1.5}$ &   &  & $\Xi_{c}(3080)^{0}$ & 3079.9$\pm$1.4   & • \\ 
$\Lambda_{c}(2765)^{+}$ &2766.6$\pm$2.4 &   & &$\Xi_{c}(2980)^{0}$ & 2968.0$\pm$2.6  & • & • \\ 
  • & • &  &  & • & • \\ 
\bottomrule
\end{tabular}
\end{table*}

The Hypercentral Constituent Quark  Model (hCQM) with color coulomb plus power potential was used for the systematic calculations of ground state masses of various baryons \cite{bhavin,15,95,93}. The hCQM scheme  accounts to an average two-body potential for the three quark system over the hyper angle. After successful calculation of ground state masses, we are implanting this model to calculate the masses of singly charmed baryons upto higher radial and orbital excitation. In this paper, we use hyper color coulomb plus linear potential as we have used in case of spectroscopy of mesons \cite{11}. We study four singly charmed baryons $\Lambda_{c}^{+}$(\textit{udc}), $\Sigma_{c}^{0}$(\textit{ddc}), $\Xi_{c}^{0}$(\textit{dsc}) and $\Omega_{c}^{0}$(\textit{ssc}) and computed their excited states (radial and orbital) masses. As we know, in SU(3) multiplicity $\Sigma_{c}^{0}$ and $\Omega_{c}^{0}$ are symmetric sextets($6_{s}$), whereas $\Lambda_{c}^{+}$ and $\Xi_{c}^{0}$ are anti-symmetric triplets($\bar{3}_{A}$) \cite{crede}. The available experimental observed states of these baryons are mentioned in Table \ref{tab:table1}. \\

We also construct the Regge trajectories for these baryons in ($n, M^2$) plane, where one can test their linearity, parallelism and equidistant. This study produces the information about the hadron dynamics and it also important for the hadron production and high energy scattering \cite{ebert2011}. It is very much important to reproduce the baryon spectrum in model, likewise the study of other properties are also important. Hence, we calculate the magnetic moment and transition magnetic moment using effective quark masses. We also determined radiative decay widths.\\

This paper is organized as follows: The description of hypercentral Constituent Quark Model (hCQM) is given in section 2. A systematic mass spectroscopy calculation have been performed in this model for four singly charmed baryons. We analyze and discuss our results in section 3. We plotted Regge Trajectories and explained it for these baryons in section 4. The magnetic moments and radiative decay widths are also determined for these baryons in section 5. Finally, we draw conclusions in section 6.

\section{The Model}

Baryons are made of three constituent quarks. The relevant degrees of freedom for their motion are related by the Jacobi coordinates ($\vec{\rho}$ and $\vec{\lambda}$) are given by \cite{Bijker} as
\begin{subequations}
\begin{equation}
\vec{\rho} = \dfrac{1}{\sqrt{2}}(\vec{r_{1}} - \vec{r_{2}})
\end{equation}
\begin{equation}
\vec{\lambda} =\dfrac{m_1\vec{r_1}+m_2\vec{r_2}-(m_1+m_2)\vec{r_3}}{\sqrt{m_1^2+m_2^2+(m_1+m_2)^2}}
\end{equation}
\end{subequations}
Here $m_i$ and $\vec{r_i}$ (i = 1, 2, 3) denote the mass and coordinate of the i-th constituent quark. In present paper, the confining three-body potential is chosen within a string-like picture, where the quarks are connected by gluonic strings and the potential strings increases linearly with a collective radius $r_{3q}$ as mentioned in \cite{ginnani2015}. Accordingly the effective two body interactions can be written as
\begin{equation}
\sum_{i<j}V(r_{ij})=V(x)+. . . .
\end{equation}
The hyper radius $x$ is a collective coordinate in the hypercentral approximation and therefore the hypercentral potential contains the three-body effects. More descriptively given in ref.\cite{ginnani2015, M. Ferraris}. The  Hamiltonian of three body baryonic system in the hCQM is then expressed as
\begin{equation}
H=\dfrac{P_{x}^{2}}{2m} +V(x)
\end{equation}
where, $m=\frac{2 m_{\rho} m_{\lambda}}{m_{\rho} + m_{\lambda}}$, is the reduced mass and $x$ is the six dimensional radial hyper central coordinate of the three body system. The respective reduced masses are given by
\begin{subequations}
\begin{equation}
m_{\rho}=\dfrac{2 m_{1} m_{2}}{m_{1}+ m_{2}}
\end{equation}
\begin{equation}
 m_{\lambda}=\dfrac{2 m_{3} (m_{1}^2 + m_{2}^2+m_1m_2)}{(m_1+m_2)(m_{1}+ m_{2}+ m_{3})}
\end{equation}
\end{subequations}
We consider the masses of light quarks(u,d) as unequal. The constituent quark mass parameters used in our calculations are given below.
\begin{itemize}
\item $m_u$=0.338, $m_d$=0.350, $m_s$=0.500 and $m_c$=1.275. (all in GeV) 
\end{itemize} 
The angle of the Hyper spherical coordinates are given by $\Omega_{\rho}= (\theta_{\rho}, \phi_{\rho})$ and $\Omega_{\lambda}= (\theta_{\lambda}, \phi_{\lambda})$. We define hyper radius $x$ and hyper angle $\xi$ in terms of the absolute values $\rho$ and $\lambda$ of the Jacobi coordinates \cite{130,131,132},
\begin{equation}
x= \sqrt{\rho^{2} + \lambda^{2}}\,\,\,and\,\,\, \xi= arctan \left(\dfrac{\rho}{\lambda} \right)
\end{equation}
In the center of mass frame ($R_{c.m.} = 0$), the kinetic energy operator can be written as
\begin{equation}
-\frac{\hbar^2}{2m}(\bigtriangleup_{\rho} + \bigtriangleup_{\lambda})= -\frac{\hbar^2}{2m}\left(\frac{\partial^2}{\partial x^2}+\frac{5}{x}\frac{\partial}{\partial x}+\frac{L^2(\Omega)}{x^2}\right)
\end{equation}
where $L^2(\Omega)$=$L^2(\Omega_{\rho},\Omega_{\lambda},\xi)$ is the quadratic Casimir operator of the six-dimensional rotational group O(6) and its eigenfunctions are the hyperspherical harmonics, $Y_{[\gamma]l_{\rho}l_{\lambda}}(\Omega_{\rho},\Omega_{\lambda},\xi)$ satisfying the eigenvalue relation$L^2Y_{[\gamma]l_{\rho}l_{\lambda}}(\Omega_{\rho},\Omega_{\lambda},\xi)=-\gamma (\gamma +4) Y_{[\gamma]l_{\rho}l_{\lambda}}(\Omega_{\rho},\Omega_{\lambda},\xi)$. Here, $\gamma$ is the grand angular momentum quantum number. The hyperradial Schrodinger equation corresponds to the Eqn.(3) Hamiltonian can be written as,
\begin{equation}
\left[\dfrac{d^{2}}{d x^{2}} + \dfrac{5}{x} \dfrac{d}{dx} - \dfrac{\gamma(\gamma +4)}{x^{2}} \right] \Psi_{ \gamma}(x) = -2m[E- V(x)]\Psi_{ \gamma}(x)
\end{equation}
where $\Psi_{\gamma}$(x) is the hypercentral wave function and $\gamma$ is the grand angular quantum number. We consider a reduced hypercentral radial function, $\phi_{\gamma}(x) = x^{\frac{5}{2}}\Psi_{ \gamma}(x)$. Thus, six dimensional hyperradial Schrodinger equation reduces to,
\begin{equation}\label{eq:6}
\left[\dfrac{-1}{2m}\dfrac{d^{2}}{d x^{2}} + \dfrac{\frac{15}{4}+ \gamma(\gamma+4)}{2mx^{2}}+ V(x)\right]\phi_{ \gamma}(x)= E\phi_{\gamma}(x)
\end{equation}
Presently, we consider the hypercentral potential V(x) as the color coulomb plus linear potential with spin interaction
\begin{equation}\label{eq:7}
V(x) =  V^{0}(x)+V_{SD}(x)
\end{equation}
where $V^{0}(x)$ is given by
\begin{equation}
V^{(0)}(x)= \dfrac{\tau}{x}+ \beta x+ V_0
\end{equation}
Here, the hyper-Coulomb strength $\tau = -\frac{2}{3} \alpha_{s}$ where $\alpha_{s}$ corresponds to the strong running coupling constant; $\frac{2}{3}$ is the color factor for baryon, $\beta$(=0.14) corresponds to the string tension for baryons. $V_{0}$ is a constant value and was used to obtain ground state masses ($J^{P}$= $\frac{1}{2}^{+}$ and $\frac{3}{2}^{+}$)\cite{11}. The values for $\Lambda_{c}^{+}$, $\Sigma_{c}^{0}$, $\Xi_{c}^{0}$ and $\Omega_{c}^{0}$ are taken -0.575,-0.465,-0.468 and -0.411, respectively.\\ The strong running coupling constant $\alpha_{s}$ is given by,
\begin{equation}
\alpha_{s}= \dfrac{\alpha_{s}(\mu_{0})}{1+\dfrac{33-2n_{f}}{12 \pi} \alpha_{s}(\mu_{0}) ln \left(\dfrac{m_{1}+ m_{2}+ m_{3}}{\mu_{0}}\right)}
\end{equation}
\begin{table*}
\begin{center}

\caption{\label{tab:table8}Ground state masses of different Lattice-QCD results.}
\begin{tabular}{cccccccccccccc}
\toprule
Particle&$J^{P}$&Our&\cite{brown}&\cite{pacs}&\cite{alex}&\cite{lattice}&\cite{mathur}&\cite{can}\\
\hline
$\Lambda_{c}^{+}$&$\frac{1}{2}^{+}$&2.286&2.254(48)(31)&2.333(112)(10)&2.272(26)(33)&2.291&&-\\
$\Sigma_{c}^{0}$&$\frac{1}{2}^{+}$&2.452&2.474(48)(31)&2.467(39)(11)&2.445(32)&2.481&2.490\\
$\Sigma_{c}^{0}$&$\frac{3}{2}^{+}$&2.518&2.551(43)(25)&2.538(70)(11)&2.513(38)&2.559&2.538\\
$\Xi_{c}^{0}$&$\frac{1}{2}^{+}$&2.471&2.433(35)(30)&2.455(11)(20)&2.469(28)&-&2.473\\
$\Xi_{c}^{0}$&$\frac{3}{2}^{+}$&2.647&2.648(70)(11)&2.674(26)(12)&2.628(33)&2.655&2.554\\
$\Omega_{c}^{0}$&$\frac{1}{2}^{+}$&2.695&2.679(37)(20)&2.673(05)(12)&2.629(22)&2.681&2.678&2.783(13)\\
$\Omega_{c}^{0}$&$\frac{3}{2}^{+}$&2.767&2.738(05)(12)&2.755(37)(24)&2.709(26)&2.764&2.752&2.837(18)\\
\bottomrule
\end{tabular}
\end{center}
\end{table*}
\begin{table*}
\caption{\label{tab:table2} Mass Spectra of $\Lambda_{c}^{+}$ Baryon.}
\begin{tabular}{cccccccccccccccc}
\toprule
 $n^{2S+1}L_{J}$&Our&Exp.\cite{olive}&\cite{ebert2011}&\cite{chen2015}&\cite{yoshida}&\cite{Roberts2008}&\cite{yama}&\cite{valcarce}&\cite{98}&Lattice\cite{4}\\
\hline
$1^{2}S_{\frac{1}{2}}$&	2.287	&2.286$\pm$0.0014		&	2.286	&	2.286	&	2.285	&	2.268	&	2.268	&	2.285	&		&	2.280(41)	\\
$2^{2}S_{\frac{1}{2}}$&	2.758	&	2.766$\pm$0.024	&	2.769	&	2.766	&	2.857	&	2.791	&	2.791	&	2.785	&		&		\\
$3^{2}S_{\frac{1}{2}}$&	3.134	&		&	3.130	&	3.112	&	3.123	&		&	2.983	&		&		&		\\
$4^{2}S_{\frac{1}{2}}$&	3.477	&		&	3.430	&	3.397	&		&		&	3.154	&		&		&		\\
$5^{2}S_{\frac{1}{2}}$&	3.787	&		&	3.715	&		&\\	
	\hline
$(1^2P_{1/2})$&	2.694	&	2.592$\pm$0.0028	&	2.589	&	2.591	&	2.628	&	2.625	&	2.625	&	2.627	&	2.600	&	2.578(289)	\\
$(1^2P_{3/2})$&	2.640	&2.628$\pm$0.0019		&	2.627	&	2.629	&	2.630	&	2.816	&	2.830	&	2.880	&	2.650	&		\\
\hline
$(2^2P_{1/2})$&	3.062	&$2.939^{+1.4}_{-1.5}$		&	2.983	&	2.989	&	2.890	&	2.636	&		&		&		&		\\
$(2^2P_{3/2})$&	3.015	&		&	3.005	&	3.000	&	2.917	&	2.830	&		&		&		&		\\
\hline
$(3^2P_{1/2})$&	3.397	&		&	3.303	&	3.296	&	2.933	&	2.872	&		&		&		&		\\
$(3^2P_{3/2})$&	3.354	&		&	3.222	&	3.301	&	2.956	&		&		&		&		&		\\
\hline
$(4^2P_{1/2})$&	3.705	&		&	3.588	&		&		&		&		&		&		&		\\
$(4^2P_{3/2})$&	3.668	&		&	3.606	&		&		&		&		&		&		&		\\
\hline
$(5^2P_{1/2})$&	3.997	&		&	3.852	&		&		&		&		&		&		&		\\
$(5^2P_{3/2})$&	3.962	&		&	3.869	&		&		&		&		&		&		&		\\
\hline
$(1^2D_{3/2})$&	2.924	&		&	2.874	&	2.857	&		&		&	3.12	&		&		&		\\
$(1^2D_{5/2})$&	2.854	&2.881$\pm$0.0035		&	2.880	&	2.879	&	2.922	&	2.887	&	3.125	&	2.888	&	2.882	&		\\
\hline
$(2^2D_{3/2})$&	3.263	&		&	3.189	&	3.188	&		&		&	3.194	&		&		&		\\
$(2^2D_{5/2})$&	3.204	&		&	3.209	&	3.198	&	3.202	&		&	3.194	&		&		&		\\
\hline
$(1^2F_{5/2})$&	3.130	&		&	3.097	&	3.075	&		&		&	3.092	&		&		&		\\
$(1^2F_{7/2})$&	3.052	&		&	3.078	&	3.092	&		&		&	3.128	&		&		&		\\
\bottomrule
\end{tabular}
\end{table*}
We get $\alpha_{s}$=0.6 with $\mu_{0}$=1 GeV and number of flavors are 4. If we compare Eqn.(\ref{eq:6}) with the usual three dimensional radial Schrodinger equation, the resemblance between angular momentum and hyper angular momentum is given by \cite{9}, ${l(l+1)\rightarrow \frac{15}{4}+ \gamma(\gamma+4)}$. The spin-dependent part of Eqn.(9), $V_{SD}(x)$ contains three types of the interaction terms, such as the spin-spin term $V_{SS} (x)$, the spin-orbit term $V_{\gamma S}(x)$ and tensor term $V_{T}(x)$ given by \cite{12},
\begin{eqnarray}
V_{SD}(x)= V_{SS}(x)(\vec{S_{\rho}}.\vec{S_\lambda})
+ V_{\gamma S}(x) (\vec{\gamma} \cdot \vec{S})&&  \nonumber \\ + V_{T} (x)
\left[ S^2-\dfrac{3(\vec{S }\cdot \vec{x})(\vec{S} \cdot \vec{x})}{x^{2}} \right]
\end{eqnarray}
The Spin-orbit and the tensor term describe the fine structure of the states, while the spin-spin term gives the spin singlet triplet splittings. The coefficient of these spin-dependent terms of Eq$^{n}$.(10) can be written in terms of the vector, $V_{V}(x)=\frac{\tau}{x}$ and scalar, $V_{S}(x)=\beta x$ parts of the static potential as
\begin{equation}
V_{\gamma S} (x) = \dfrac{1}{2 m_{\rho} m_{\lambda}x}  \left(3\dfrac{dV_{V}}{dx} -\dfrac{dV_{S}}{dx} \right)
\end{equation}
\begin{equation}
V_{T}(x)=\dfrac{1}{6 m_{\rho} m_{\lambda}} \left(3\dfrac{d^{2}V_{V}}{dx^{2}} -\dfrac{1}{x}\dfrac{dV_{V}}{dx} \right)
\end{equation}

\begin{equation}
V_{SS}(x)= \dfrac{1}{3 m_{\rho} m_{\lambda}} \bigtriangledown^{2} V_{V}
\end{equation}
Instead of the six dimensional delta function which appear into spin-spin interaction term of $Eq^n.$(15), we use smear function similar to the one given by \cite{15,95}
\begin{equation}
V_{SS}(x)= \dfrac{-A}{6 m_{\rho} m_{\lambda}}  \frac{e^{-x/x_0}}{x x_0^2}
\end{equation}
where $x_{0}$ is the hyperfine parameter of the model. We take $A ={A_{0}}/{(n + \gamma +\frac{3}{2})^{2}} $ , where $A_{0}$ is arbitary constant. The baryon spin average mass in this hypercentral model is given by $M_B = \sum_{i=1} m_{i} + BE$. We numerically solve the six dimensional Schrodinger equation using Mathematica notebook \cite{lucha}. We have followed the $^{(2S+1)} {\gamma}_{J}$ notations for spectra of baryons.

\section{Mass Spectra:Discussions}
We have calculated the masses of radial and orbital excited heavy baryons $\Lambda_{c}^{+}$, $\Sigma_{c}^{0}$, $\Xi_{c}^{0}$ and $\Omega_{c}^{0}$. Our masses of ground states are compared with different Lattice QCD models \cite{4,brown,alex,pacs,lattice,mathur,can} in Table 2. Refs.\cite{Migura, zahra} has also performed ground state calculations in their papers. The obtained masses of radial excited states charm baryons with $J = \frac{1}{2}$, $J = \frac{3}{2}$ and $J =\frac{5}{2}$(positive parity) and the masses of negative parity states baryons with $J = \frac{1}{2}$, $J = \frac{3}{2}$, $J = \frac{5}{2}$ and $J = \frac{7}{2}$ are listed in Table \ref{tab:table2}-\ref{tab:table5}. \\ 

Until now, only Ref. \cite{ebert2011} focused on the mass spectra of the radial as well as orbital excited heavy baryons; precisely for 2S-5S and 1P-5P, 1D-2D and 1F states with a use of the relativistic quark potential model in the quark-diquark picture. Ref. \cite{chen2015} has calculated higher excited states of $\Lambda_c^{+}$ (upto 3S, 3P, 1D, 1F and 1G states) and $\Xi_c$ (upto 4S and 3P, 1D, 1F and 1G states) within the relativistic flux tube (RFT) model. A non-relativistic quark model with harmonic oscillator potential also showed excited mass spectra of $\Lambda_c$, $\Sigma_{c}$ and $\Omega_{c}$ upto 3S, 3P and 3D states \cite{yoshida}. The excited mass specta in HQS limit is calculted by \cite{yama} for $\Lambda_c$(2S-4S, 1P, 1D and 1F), $\Sigma_{c}$ and $\Xi_{c}$(2S-3S, 1P and 1D). Ref. \cite{Roberts2008, valcarce,4} has calculated 1S and 1P states whereas \cite{98} calculated only 1P state for all these baryons. All these models are compared with our obtained results in respective tables of the baryons. Several S-wave, P-wave and D-wave single charm baryon masses are given in Table \ref{tab:table1} with their known experimental masses. We anticipated heavy baryons having low lying states with $J^{P}$, $\frac{1}{2}^{+}$ and $\frac{3}{2}^{+}$ and the higher states with $J^{P}$, $\frac{5}{2}^{+}$, $\frac{7}{2}^{+}$,$\frac{1}{2}^{-}$,$\frac{3}{2}^{-}$,$\frac{5}{2}^{-}$,$\frac{7}{2}^{-}$ and $\frac{9}{2}^{-}$ in accordance to S= $\frac{1}{2}$ and $\frac{3}{2}$. The mass spectra of $\Lambda_{c}^{+}$, $\Sigma_{c}^{0}$, $\Xi_{c}^{0}$ and $\Omega_{c}^{0}$ Baryons are given in following subsection 3.1-3.4.  

\begin{table*}
\begin{center}
\tabcaption{\label{tab:table3} Mass spectra of $\Sigma_{c}^{0}$ baryon.}
\resizebox{\columnwidth}{!}{%
\begin{tabular}{c|cccccccccccccccc}
\toprule
$n^{2S+1}L_{J}$&Our&Exp.\cite{olive}&\cite{ebert2011}&\cite{Roberts2008}&\cite{yoshida}&\cite{yama}&\cite{valcarce}&\cite{98}&Lattice \cite{4}\\
\hline
$1^{2}S_{\frac{1}{2}}$&	2.452	&2.453$\pm$0.0016		&	2.443	&	2.455	&	2.460	&	2.455	&	2.435	&		&	2.434(46)	\\
$2^{2}S_{\frac{1}{2}}$&	2.891	&		&	2.901	&	2.958	&	3.029	&	2.958	&	2.904	&		&		\\
$3^{2}S_{\frac{1}{2}}$&	3.261	&		&	3.271	&		&	3.271	&	3.115	&		&		&		\\
$4^{2}S_{\frac{1}{2}}$&	3.593	&		&	3.581	&		&		&		&		&		&		\\
$5^{2}S_{\frac{1}{2}}$&	3.900	&		&	3.861	&		&		&		&		&		&		\\
\hline
$1^{4}S_{\frac{3}{2}}$&	2.518	&2.518$\pm$0.0006		&	2.519	&	2.519	&	2.523	&	2.519	&	2.502	&		&	3.713(16)	\\
$2^{4}S_{\frac{3}{2}}$&	2.917	&		&	2.936	&	2.995	&	3.065	&	2.995	&	2.944	&		&		\\
$3^{4}S_{\frac{3}{2}}$&	3.274	&		&	3.293	&		&	3.094	&	3.116	&		&		&		\\
$4^{4}S_{\frac{3}{2}}$&	3.601	&		&	3.598	&		&		&		&		&		&		\\
$5^{4}S_{\frac{3}{2}}$&	3.906	&		&	3.873	&		&		&		&		&		&		\\
\hline
$(1^2P_{1/2})$&	2.809	&	2.806$\pm$0.0018	&	2.799	&	2.748	&	2.802	&	2.848	&	2.772	&	2.730	&	3.785(16)	\\
$(1^2P_{3/2})$&	2.755	&		&	2.798	&	2.763	&	2.807	&	2.763	&	2.772	&	2.800	&	2.740(99)	\\
$(1^4P_{1/2})$&	2.835	&		&	2.713	&	2.768	&		&		&		&		&		\\
$(1^4P_{3/2})$&	2.782	&		&	2.773	&	2.776	&		&		&		&		&		\\
$(1^4P_{5/2})$&	2.710	&		&	2.789	&	2.790	&	2.839	&	2.790	&		&	2.890	&		\\
\hline
$(2^2P_{1/2})$&	3.174	&		&	3.172	&		&	2.826	&		&	2.893	&		&		\\
$(2^2P_{3/2})$&	3.128	&		&	3.172	&		&	2.837	&		&		&		&		\\
$(2^4P_{1/2})$&	3.196	&		&	3.125	&		&		&		&		&		&		\\
$(2^4P_{3/2})$&	3.151	&		&	3.151	&		&		&		&		&		&		\\
$(2^4P_{5/2})$&	3.090	&		&	3.161	&		&	3.316	&		&		&		&		\\
\hline
$(3^2P_{1/2})$&	3.505	&		&	3.488	&		&	2.909	&		&		&		&		\\
$(3^2P_{3/2})$&	3.465	&		&	3.486	&		&	2.91	&		&		&		&		\\
$(3^4P_{1/2})$&	3.525	&		&	3.455	&		&		&		&		&		&		\\
$(3^4P_{3/2})$&	3.485	&		&	3.469	&		&		&		&		&		&		\\
$(3^4P_{5/2})$&	3.433	&		&	3.475	&		&	3.521	&		&		&		&		\\
\hline
$(4^2P_{1/2})$&	3.814	&		&	3.770	&		&		&		&		&		&		\\
$(4^2P_{3/2})$&	3.777	&		&	3.768	&		&		&		&		&		&		\\
$(4^4P_{1/2})$&	3.832	&		&	3.743	&		&		&		&		&		&		\\
$(4^4P_{3/2})$&	3.796	&		&	3.753	&		&		&		&		&		&		\\
$(4^4P_{5/2})$&	3.747	&		&	3.757	&		&		&		&		&		&		\\
\hline
$(1^4D_{1/2})$&	3.036	&		&	3.041	&		&		&		&		&		&		\\
$(1^2D_{3/2})$&	3.112	&		&	3.043	&		&		&	3.095	&		&		&		\\
$(1^4D_{3/2})$&3.061		&		&	3.040	&		&		&		&		&		&		\\
$(1^2D_{5/2})$&	2.993	&		&	3.038	&3.003		&	3.099	&	3.003	&		&		&		\\
$(1^4D_{5/2})$&	2.968	&		&	3.023	&		&		&		&		&		&		\\
$(1^4D_{7/2})$&	2.909	&		&	3.013	&		&		&	3.015	&		&		&		\\
\hline
$(2^4D_{1/2})$&	3.376	&		&	3.370	&		&		&		&		&		&		\\
$(2^2D_{3/2})$&	3.398	&		&	3.366	&		&		&		&		&		&		\\
$(2^4D_{3/2})$&	3.442	&		&	3.364	&		&		&		&		&		&		\\
$(2^2D_{5/2})$&	3.316		&		&	3.365	&		&3.114		&		&		&		&		\\
$(2^4D_{5/2})$&	3.339&		&	3.349	&		&		&		&		&		&		\\
$(2^4D_{7/2})$&	3.265	&		&	3.342	&		&		&		&		&		&		\\
\hline
$(1^4F_{3/2})$&	3.332	&		&	3.288	&		&		&		&		&		&		\\
$(1^2F_{5/2})$&	3.245	&		&	3.283	&		&		&		&		&		&		\\
$(1^4F_{5/2})$&	3.268	&		&	3.254	&		&		&		&		&		&		\\
$(1^4F_{7/2})$&	3.189	&		&	3.253	&		&		&		&		&		&		\\
$(1^2F_{7/2})$&	3.165	&		&	3.227	&		&		&		&		&		&		\\
$(1^4F_{9/2})$&	3.094	&		&	3.209	&		&		&		&		&		&		\\
\bottomrule
\end{tabular}%
}
\end{center}
\end{table*}

\begin{table*}
\begin{center}
\caption{\label{tab:table4} Mass spectra of of $\Xi_{c}^{0}$ baryon.}
\resizebox{\columnwidth}{!}{%
\begin{tabular}{c|cccccccccccccccc}
\toprule
$n^{2S+1}L_{J}$&Our&Exp.\cite{olive}&\cite{ebert2011}&\cite{chen2015}&\cite{yama}&\cite{valcarce}&\cite{98}&Lattice \cite{4}\\
\hline
$1^{2}S_{\frac{1}{2}}$&	2.471	&2.470$^{+0.34}_{-0.80}$		&	2.476	&	2.467	&	2.466	&	2.471	&		&	2.442(31)	\\
$2^{2}S_{\frac{1}{2}}$&	2.937	&	2.968$\pm$ 0.0026		&	2.959	&	2.959	&	2.924	&		&		&		\\
$3^{2}S_{\frac{1}{2}}$&	3.303	&		&	3.323	&	3.325	&	3.183	&		&		&		\\
$4^{2}S_{\frac{1}{2}}$&	3.626	&		&	3.642	&	3.629	&		&		&		&		\\
$5^{2}S_{\frac{1}{2}}$&	3.921	&		&	3.909	&		&		&		&		&		\\
\hline
$1^{4}S_{\frac{3}{2}}$&	2.647	&2.645$\pm$ 0.0005	&		&		&		&	2.642	&		&2.608(35) 		\\
$2^{4}S_{\frac{3}{2}}$&	3.004	&		&		&		&		&		&		&		\\
$3^{4}S_{\frac{3}{2}}$&	3.338	&		&		&		&		&		&		&		\\
$4^{4}S_{\frac{3}{2}}$&	3.646	&		&		&		&		&		&		&		\\
$5^{4}S_{\frac{3}{2}}$&	3.934	&		&		&		&		&		&		&		\\
\hline
$(1^2P_{1/2})$&	2.877	&2.791$\pm$ 0.033			&	2.792	&	2.779	&	2.773	&	2.799	&	2.790&	2.761(156)	\\
$(1^2P_{3/2})$&	2.834	&2.819$\pm$ 0.012			&	2.819	&	2.814	&	2.783	&	2.902	&2.830		&	2.891(68)	\\
$(1^4P_{1/2})$&	2.899	&	2.931$\pm$0.0008	&		&		&		&		&		&		\\
$(1^4P_{3/2})$&	2.856	&		&		&		&		&		&		&		\\
$(1^4P_{5/2})$&	2.798	&		&		&		&		&		&		&		\\
\hline
$(2^2P_{1/2})$&	3.222	&		&	3.179	&	3.195	&		&		&		&		\\
$(2^2P_{3/2})$&	3.189	&		&	3.201	&	3.204	&		&		&		&		\\
$(2^4P_{1/2})$&	3.239	&		&		&		&		&		&		&		\\
$(2^4P_{3/2})$&	3.206	&		&		&		&		&		&		&		\\
$(2^2P_{5/2})$&	3.162	&		&		&		&		&		&		&		\\
\hline
$(3^2P_{1/2})$&	3.544	&		&	3.5	&	3.521	&		&		&		&		\\
$(3^2P_{3/2})$&	3.512	&		&	3.519	&	3.525	&		&		&		&		\\
$(3^4P_{1/2})$&	3.561	&		&		&		&		&		&		&		\\
$(3^4P_{3/2})$&	3.528	&		&		&		&		&		&		&		\\
$(3^4P_{5/2})$&	3.484	&		&		&		&		&		&		&		\\
\hline
$(4^2P_{1/2})$&	3.837	&		&	3.785	&		&		&		&		&		\\
$(4^2P_{3/2})$&	3.808	&		&	3.804	&		&		&		&		&		\\
$(4^4P_{1/2})$&	3.851	&		&		&		&		&		&		&		\\
$(4^4P_{3/2})$&	3.823	&		&		&		&		&		&		&		\\
$(4^4P_{5/2})$&	3.784	&		&		&		&		&		&		&		\\
\hline
$(1^4D_{1/2})$&	3.147	&		&		&		&		&		&		&		\\
$(1^2D_{3/2})$&	3.109	&		&	3.059	&	3.055	&		&		&		&		\\
$(1^4D_{3/2})$&	3.090	&		&		&		&		&		&		&		\\
$(1^2D_{5/2})$&	3.058	&3.079$\pm$ 0.0017		&	3.076	&	3.076		&	3.049	&	3.071	&		\\
$(1^4D_{5/2})$&	3.039	&		&		&		&		&		&		&		\\
$(1^4D_{7/2})$&	2.995	&		&		&		&		&		&		&		\\
\hline
$(2^4D_{1/2})$&3.470	&		&		&		&		&		&		&		\\
$(2^2D_{3/2})$&	3.417	&		&	3.388	&	3.407	&		&		&		&		\\
$(2^4D_{3/2})$&3.434		&		&		&		&		&		&		&		\\
$(2^2D_{5/2})$&3.701		&		&	3.407	&	3.416	&		&		&		&		\\
$(2^4D_{5/2})$&	3.388	&		&		&		&		&		&		&		\\
$(2^4D_{7/2})$&3.330		&		&		&		&		&		&		&		\\
\hline
$(1^4F_{3/2})$&	3.360	&		&		&		&		&		&		&		\\
$(1^2F_{5/2})$&	3.291	&		&	3.278	&3.286			&	&		&		&		\\
$(1^4F_{5/2})$&	3.310	&		&		&		&		&		&		&		\\
$(1^4F_{7/2})$&	3.247	&		&		&		&		&		&		&		\\
$(1^2F_{7/2})$&	3.228	&		&	3.292	&	3.301	&		&		&		&		\\
$(1^4F_{9/2})$&	3.172	&		&		&		&		&		&		&		\\
\bottomrule
\end{tabular}%
}
\end{center}
\end{table*}

\begin{table*}
\begin{center}
\caption{\label{tab:table5} Mass Spectra of $\Omega_{c}^{0}$ baryon.}
\resizebox{\columnwidth}{!}{%
\begin{tabular}{c|ccccccccccccc}
\toprule
$n^{2S+1}L_{J}$&Our&Exp.\cite{olive}&\cite{ebert2011}&\cite{Roberts2008}&\cite{yoshida}&\cite{98}&\cite{valcarce}&\cite{4}&\cite{yama}\\
\hline
$1^{2}S_{\frac{1}{2}}$&	2.695	&2.695$\pm$0.0017	&	2.698	&	2.718	&	2.731	&		&	2.699&2.648(28)&2.718	\\
$2^{2}S_{\frac{1}{2}}$&	3.100	&		&	3.088	&	3.152	&	3.227	&		&	3.159&2.709(32)&3.152	\\
$3^{2}S_{\frac{1}{2}}$&	3.436	&		&	3.489	&		&	3.292	&		&	&&3.275	\\
$4^{2}S_{\frac{1}{2}}$&	3.737	&		&	3.814	&		&	3.814	&		&	&&3.299	\\
$5^{2}S_{\frac{1}{2}}$&	4.015	&		&	4.102	&		&	4.102	&		&	&	\\
\hline
$1^{4}S_{\frac{3}{2}}$&	2.767	&2.766$\pm$0.0002		&	2.768	&	2.776	&	2.779	&		&	2.767	\\
$2^{4}S_{\frac{3}{2}}$&	3.126	&		&	3.123	&	3.190	&	3.257	&		&	3.202	\\
$3^{4}S_{\frac{3}{2}}$&	3.450	&		&	3.51	&		&	3.285	&		&		\\
$4^{4}S_{\frac{3}{2}}$&	3.745	&		&	3.83	&		&	3.83	&		&		\\
$5^{4}S_{\frac{3}{2}}$&	4.021	&		&	4.114	&		&	4.114	&		&		\\
\hline											$(1^2P_{1/2})$ &	3.011	&		&	3.055	&	2.977	&	3.030	&	3.250	&	2.980&2.995(46)&3.046	\\
$(1^2P_{3/2})$ &	2.976	&		&	3.054	&	2.986	&	3.033	&	3.260	&	2.980&3.016(69)	&2.986\\
$(1^4P_{1/2})$ &	3.028	&		&	2.966	&	2.977	&		&		&	3.035	\\
$(1^4P_{3/2})$ &	2.993	&		&	3.029	&	2.959	&		&		&		\\
$(1^4P_{5/2})$ &	2.947	&		&	3.051	&	3.014	&	3.057	&	3.320	&&&3.014		\\
\hline
$(2^2P_{1/2})$&	3.345	&		&	3.435	&		&	3.048	&		&	3.125	\\
$(2^2P_{3/2})$&	3.315	&		&	3.433	&		&	3.056	&		&		\\
$(2^4P_{1/2})$&	3.359	&		&	3.384	&		&		&		&		\\
$(2^4P_{3/2})$&	3.330	&		&	3.415	&		&		&		&		\\
$(2^4P_{5/2})$&	3.290	&		&	3.427	&		&	3.477	&		&		\\
\hline
$(3^2P_{1/2})$&	3.644	&		&	3.754	&		&	3.048	&		&		\\
$(3^2P_{3/2})$&	3.620	&		&	3.752	&		&	3.056	&		&		\\
$(3^4P_{1/2})$&	3.656	&		&	3.717	&		&		&		&		\\
$(3^4P_{3/2})$&	3.632	&		&	3.737	&		&		&		&		\\
$(3^4P_{5/2})$&	3.601	&		&	3.744	&		&	3.620	&		&		\\
\hline
$(4^2P_{1/2})$&	3.926	&		&	4.037	&		&		&		&		\\
$(4^2P_{3/2})$&	3.903	&		&	4.036	&		&		&		&		\\
$(4^4P_{1/2})$&	3.938	&		&	4.009	&		&		&		&		\\
$(4^4P_{3/2})$&	3.915	&		&	4.023	&		&		&		&		\\
$(4^4P_{5/2})$&	3.884	&		&	4.028	&		&		&		&		\\
\hline
$(1^4D_{1/2})$&	3.215	&		&	3.287	&		&		&		&		\\
$(1^2D_{3/2})$&	3.231	&		&	3.298	&		&		&		&		\\
$(1^4D_{3/2})$&	3.262	&		&	3.282	&		&		&		&		\\
$(1^2D_{5/2})$&	3.188	&		&	3.297	&	3.196	&	3.288	&		&	&&3.273	\\
$(1^4D_{5/2})$&	3.173	&		&	3.286	&		&		&		&		\\
$(1^4D_{7/2})$&	3.136	&		&	3.283	&		&		&		&		\\
\hline
$(2^4D_{1/2})$&	3.524	&		&	3.623	&		&		&		&		\\
$(2^2D_{3/2})$&	3.538	&		&	3.627	&		&		&		&		\\
$(2^4D_{3/2})$&	3.565	&		&	3.613	&		&		&		&		\\
$(2^2D_{5/2})$&	3.502	&		&	3.626	&		&		&		&		\\
$(2^4D_{5/2})$&	3.488	&		&	3.614	&		&		&		&		\\
$(2^4D_{7/2})$&	3.456	&		&	3.611	&		&		&		&		\\
\hline
$(1^4F_{3/2})$&	3.457	&		&	3.533	&		&		&		&		\\
$(1^2F_{5/2})$&	3.403	&		&	3.522	&		&		&		&		\\
$(1^4F_{5/2})$&	3.418	&		&	3.515	&		&		&		&		\\
$(1^4F_{7/2})$&	3.369	&		&	3.54	&		&		&		&		\\
$(1^2F_{7/2})$&	3.354	&		&	3.498	&		&		&		&		\\
$(1^4F_{9/2})$&	3.310	&		&	3.485	&		&		&		&		\\
\bottomrule
\end{tabular}%
}
\end{center}
\end{table*}

\subsection{$\Lambda_c^{+}$ Baryon}

$\Lambda_{c}^{+}$ was the first singly charmed baryon has been discovered at BNL \cite{1} in 1975. The experimental known-unknown states are tabulated in Table \ref{tab:table1}. The experimental known ground state is $\Lambda_{c}(2286)^{+}$. The first orbital excited states are $\Lambda_{c}(2595)^{+}$ and $\Lambda_{c}(2625)^{+}$ with quantum numbers $\Lambda_{c}^{+}(\frac{1}{2}^{-})$ and $\Lambda_{c}^{* +}(\frac{3}{2}^{-})$ respectively. $\Lambda_{c}(2880)^{+}$ with $J^P=\frac{5}{2}^{+}$ is also known. Two more states namely $\Lambda_{c}(2765)^{+}$ and $\Lambda_{c}(2940)^{+}$ are observed by CLEO collaboration \cite{cleo} and Babar collaboration \cite{3} but their $J^{P}$ values are still unknown. We have calculated the mass spectra of $\Lambda_{c}^{+}$ baryon for 1S-5S, 1P-5P, 1D-2D and 1F states (see Table \ref{tab:table2}).\\ 

Our ground state is close to Lattice QCD results (see Table \ref{tab:table8}) and other theoretical predictions. The radial excited states 2S-5S show ($\approx$4-78 MeV) difference with Ref. \cite{ebert2011}. The comparison with other models are also mentioned. We obtained 2.758 GeV in 2S state which is near to Ref. \cite{ebert2011, chen2015, olive}. Thus, we assigned $J^P = \frac{1}{2}^{+}$ to $\Lambda_{c}(2765)^{+}$. The first orbital $\Lambda_{c}^{+}(\frac{1}{2}^{-})$ state value is 102 MeV less than our result, but Ref. \cite{Roberts2008,yoshida,yama,valcarce} are closer while $\Lambda_{c}^{* +}(\frac{3}{2}^{-})$ state differnce value is only 12 MeV. Our 2P state value is quite higher then other prediction, though we would like to assign $\Lambda_{c}^{* +}(\frac{3}{2}^{-})$ as 2P state with 123 MeV difference. Ebert et al.\cite{ebert2011}, Chen et al.\cite{chen2015} and Yamaguchi et al.\cite{yama} have calculated D($\frac{3}{2}^{+},\frac{5}{2}^{+}$) and F state($\frac{5}{2}^{-},\frac{7}{2}^{-}$) and our values are higher than them. Our 1D state ($\frac{5}{2}^{+}$) is also closer to the experimental value(27 Mev difference).

\subsection{$\Sigma_c^0$ Baryon}

PDG(2014) has listed two ground states $\Sigma_{c}(2455)^{0}$ and $\Sigma_{c}(2520)^{0}$ with their $J^P$ values $\frac{1}{2}^{+}$ and $\frac{3}{2}^{+}$. $\Sigma_{c}(2800)$ is also found but the quantum number is undefined. We have calculated the mass spectroscopy of $\Sigma_{c}^{+}$ baryon for 1S-5S, 1P-4P, 1D-2D and 1F states (see Table \ref{tab:table3}).\\

The ground states are closed with Lattice-QCD results(see Table \ref{tab:table8}) as well as experimental results. The radial excited states 2S-5S show ($\approx$ 10-39 MeV) difference for $\Sigma_{c}$ and ($\approx$ 19-33 MeV) difference for $\Sigma_{c}^*$ with Ref.\cite{ebert2011}. These results are also close to the \cite{yoshida,yama}. The $J^{P}$ value of $\Sigma_{c}(2800)$ state is unknown experimentally. We compare, our 1P state with $J^P$($\frac{1}{2}^{-}$, $\frac{3}{2}^{-}$ and $\frac{5}{2}^{-}$) and it shows 3, 51, 96 MeV difference respectively. These states are also calculated by \cite{ebert2011, Roberts2008, yoshida, yama,98,4} and are reasonably close. Ref. \cite{Roberts2008,yama} shows only 35 MeV difference and \cite{ebert2011} shows 7 MeV difference with our 1D ($\frac{5}{2}^{+}$) state. Ref. \cite{87} has also calculated for $J^P$($\frac{1}{2}^{-}$ and $\frac{3}{2}^{-}$) and obtained 2.74$\pm$0.20 GeV and Ref. \cite{92} has also calculated m($\Sigma_{c}$)= 2.84 $\pm$ 0.11 GeV for $J^{P}=\frac{3}{2}^{-}$.
\subsection{$\Xi_c^0$ Baryon}

$\Xi_{c}$ baryon was announced by Belle \cite{2} and also by BABAR collaboration \cite{3}. $\Xi_{c}(2470)$ and $\Xi_{c}^{*}(2645)$ are two lowest-lying ground states. First orbital excited state $\Xi_{c}(2790)$ with $J^{P}= \frac{1}{2}^{-}$ and $\Xi_{c}^{*}(2815)$ with $J^{P}= \frac{3}{2}^{-}$ are well established experimentally \cite{olive}. 1D state ($\frac{5}{2}^{+}$) $\Xi_{c}^{0}(3080)$ is also experimentally known.
$\Xi_{c}(2930)$ is only observed by BABAR \cite{3} 
and $\Xi_{c}(2980)^{0}$ is found by Belle Collaboration \cite{belle}. Although these baryons are well established experimentally, its quantum numbers are still unknown. We have calculated the mass spectroscopy of $\Xi_{c}^{0}$ baryon for 1S-5S, 1P-4P, 1D-2D and 1F states (see Table \ref{tab:table4}).\\

The ground state $\Xi_c^0$ is in accordance with different phenomenological models and Lattice-QCD results (Table \ref{tab:table8}). $\Xi_c^{*0}(\frac{3}{2}^{+}$) results are perfectly matched with Refs. \cite{olive,brown,valcarce}. Our 2S state mass is 31 MeV lower than $\Xi_{c}(2980)^{0}$. Other theoretical results \cite{ebert2011, chen2015, yama} are also close to experimental result. So, that it could be 2S($\frac{1}{2}^{+}$) state. Our obtained 1P state shows higher value than Refs. \cite{olive,4,ebert2011,yoshida,chen2015,yama,98,valcarce} while the other 1P state with $J^{P}(\frac{3}{2}^{-}$) is closer to experimental and all above models. We compare our $\Xi_c^0(\frac{1}{2}^{-}$) with one more $\Xi_{c}(2930)$ state. We Predict this unknown state as 1P as the difference of the values are just 32 MeV. 1D state mass is 21 MeV lower than experimental value. Rest of the states(2D and 1F) with S=$\frac{1}{2}$ are also reasonably close with Refs. \cite{ebert2011, chen2015}. 

\subsection{$\Omega_c^{0}$ Baryon}
 The WA62 collaboration announced the first ground state of $\Omega_{c}$ \cite{WA62}. This baryon is least known experimentally. Only two ground states $\Omega_{c}(2695)$ and $\Omega_{c}(2770)$ are observed yet. Though we have calculated the mass spectroscopy of $\Omega_{c}^{0}$ baryon for 1S-5S, 1P-4P, 1D-2D and 1F states (see Table \ref{tab:table5}) like other baryons. 

Our ground states $\Omega_{c}^{0}$ and $\Omega_{c}^{*0}$ are in good agreement with other theoretical predictions as well as experimental measurements. and Lattice-QCD results. We mainly compare our results with D.ebert et al.\cite{ebert2011} and Yoshida et al., \cite{yoshida}. As we move to higher excited states we can observe that the difference of masses with other model get increases. Though our obtained masses are lower than them. Our results are ($\approx$ 100 MeV close to the W. Roberts calculation \cite{Roberts2008}.

\section{Regge Trajectories}
The calculted masses are used to plot Regge trajectrories for $\Lambda_{c}^{+}$, $\Sigma_{c}^{0}$, $\Xi_{c}^{0}$ and $\Omega_{c}^{0}$ in $M^{2} \rightarrow$ n plane. The output of the graphs show good agreement with experimental measurements. We use (n, $M^{2}$) Regge trajectories 
\begin{equation}
n=\beta M^2+ \beta_{0}
\end{equation}
\noindent Where, $\beta$ and $\beta_{0}$ are slope and intercept, respectively. n=\textbf{n}-1, \textbf{n} is principal quantum number. The values of $\beta$ and $\beta_0$ are shown in table~\ref{tab:table7} for the different baryons. As describe in previous section, we have calculated masses of S, P and D states which are used to construct Regge trajectories in (n, $M^{2}$) plane. The ground and radial excited states S with $J^{P}=\frac{1}{2}^{+}$ and the orbital excited state P with $J^{P}= \frac{1}{2}^{-}$, D with $J^{P}= \frac{5}{2}^{+}$ are plotted. These trajectories for different baryons are presented in Figs.[1-2]. \\

 $\Lambda_c^{+}$ and $\Xi_c^{0}$ are presented in Fig.[1] and $\Sigma_c^0$ and $\Omega_c^0$ are included in Fig.[2]. Our obtained results upto (L=2) are shown with experimental states. These known charmed baryons are mentioned with their names in all figures. Straight lines were obtained by the linear fitting in both figures. We observe that the square of the calculated masses fit very well to the linear trajectory and almost parallel and equidistant in S, P, D states. We can determine the possible quantum numbers and prescribe them to particular Regge trajectory with the help of our obtained results.\\
\begin{center}
\tabcaption{Fitted slope($\beta$) and intercept($\beta_{0}$) of the regge trajectories}
\label{tab:7}
\begin{tabular*}{80mm}{c@{\extracolsep{\fill}}cccc}
\toprule
Baryon&$J^{P}$ &State& $\beta$&$\beta_{0}$\\
 \hline
$\Lambda_{c}$&$\frac{1}{2}^{+}$&S&0.440$\pm$0.0029&-2.327$\pm$0.030\\
$\Lambda_{c}$&$\frac{1}{2}^{-}$&P&0.459$\pm$0.003&-3.314$\pm$0.036\\
$\Lambda_{c}$&$\frac{5}{2}^{+}$&D&0.472
&-3.841\\
\hline
$\Sigma_{c}$&$\frac{1}{2}^{+}$&S&0.436$\pm$0.001&-2.628$\pm$0.015\\
$\Sigma_{c}$&$\frac{1}{2}^{-}$&P&0.451$\pm$0.003&-3.548$\pm$0.029\\
$\Sigma_{c}$&$\frac{5}{2}^{+}$&D&0.459&-4.230\\
\hline
$\Xi_{c}$&$\frac{1}{2}^{+}$&S&0.4335$\pm$0.006&-2.696$\pm$0.073\\
$\Xi_{c}$&$\frac{1}{2}^{-}$&P&0.465$\pm$0.002&-3.841$\pm$0.028\\
$\Xi_{c}$&$\frac{5}{2}^{+}$&D&0.497&-4.651\\
\hline
$\Omega_{c}$&$\frac{1}{2}^{+}$&S&0.453$\pm$0.005&-3.320$\pm$0.054\\
$\Omega_{c}$&$\frac{1}{2}^{-}$&P&0.473$\pm$0.001&-4.285$\pm$0.017\\
$\Omega_{c}$&$\frac{5}{2}^{+}$&D&0.476&-4.790\\
\hline
\end{tabular*}
\end{center}
\begin{figure*}
\centering
\begin{minipage}[b]{0.40\linewidth}
\includegraphics[scale=0.30]{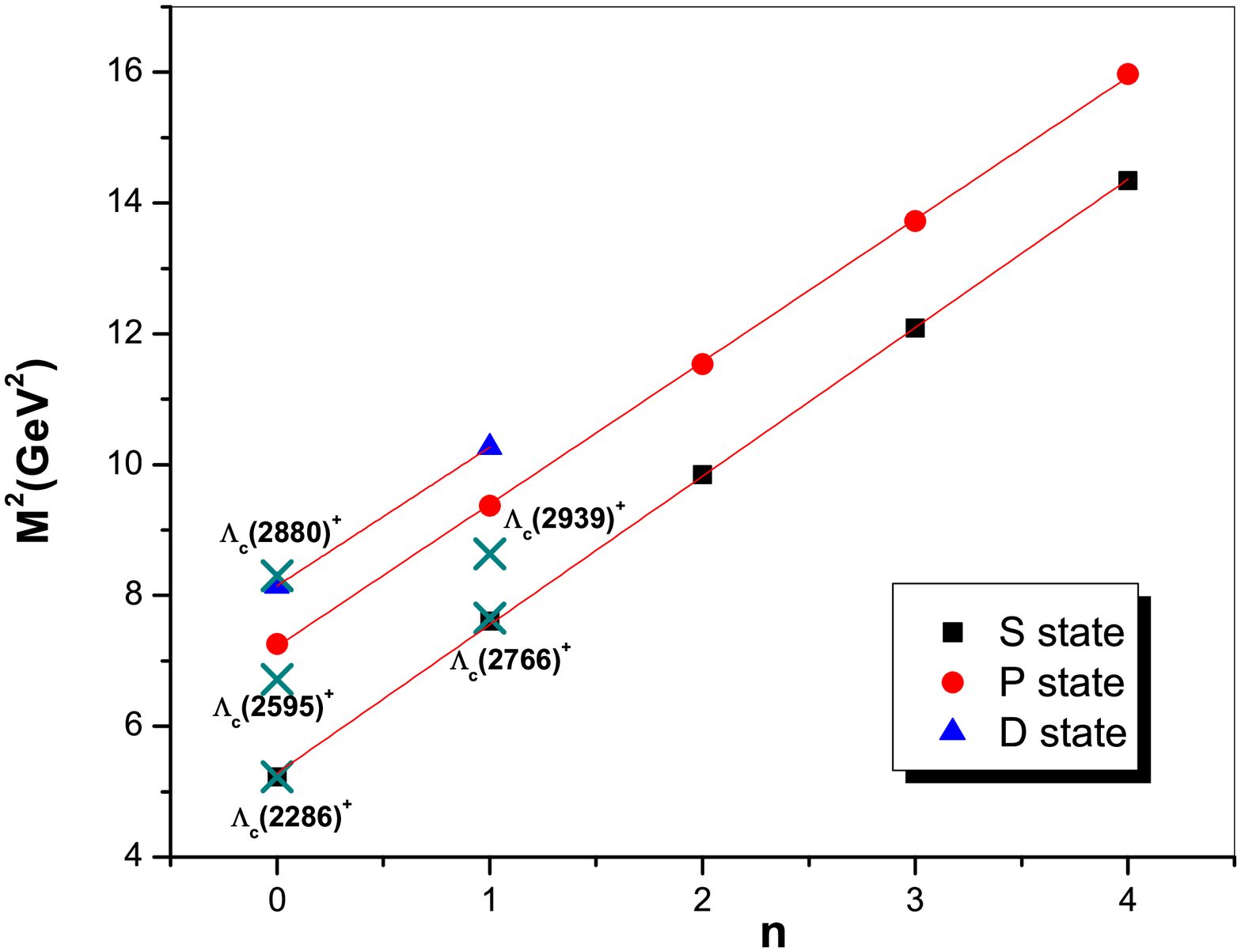}
\label{fig:10}
\end{minipage}
\quad
\begin{minipage}[b]{0.40\linewidth}
\includegraphics[scale=0.30]{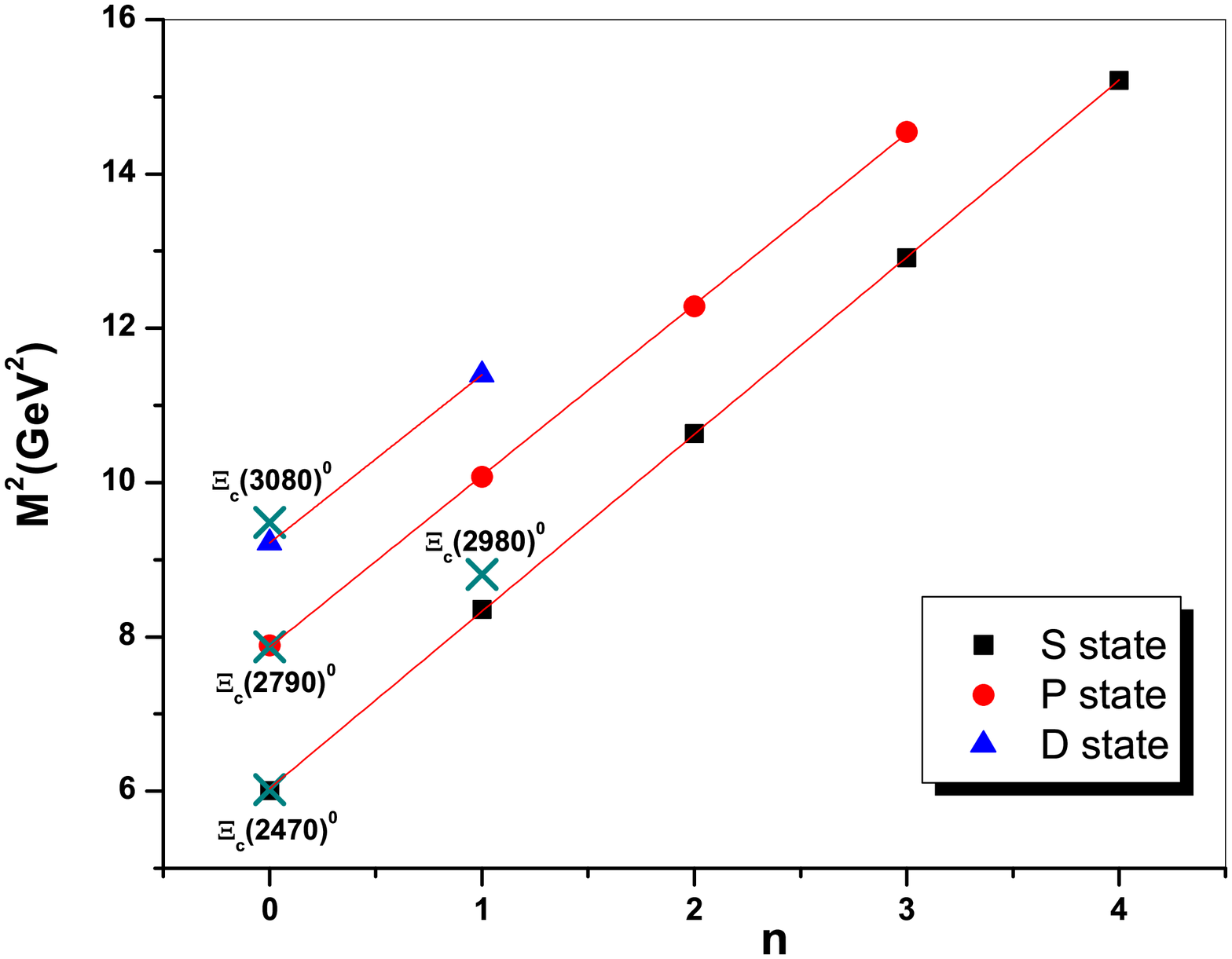}
\label{fig:10}
\end{minipage}
\caption{\label{fig:epsart} Regge Trajectory of $\Lambda_{c}^{0}$(left) and
$\Xi_{c}^{+}$(right) for $J^{P}$ values $\frac{1}{2}^{+}$,$\frac{1}{2}^{-}$ and $\frac{5}{2}^{+}$ with different states. Available Expt. states are given with particle name.}
\end{figure*}

\begin{figure*}
\centering
\begin{minipage}[b]{0.40\linewidth}
\includegraphics[scale=0.30]{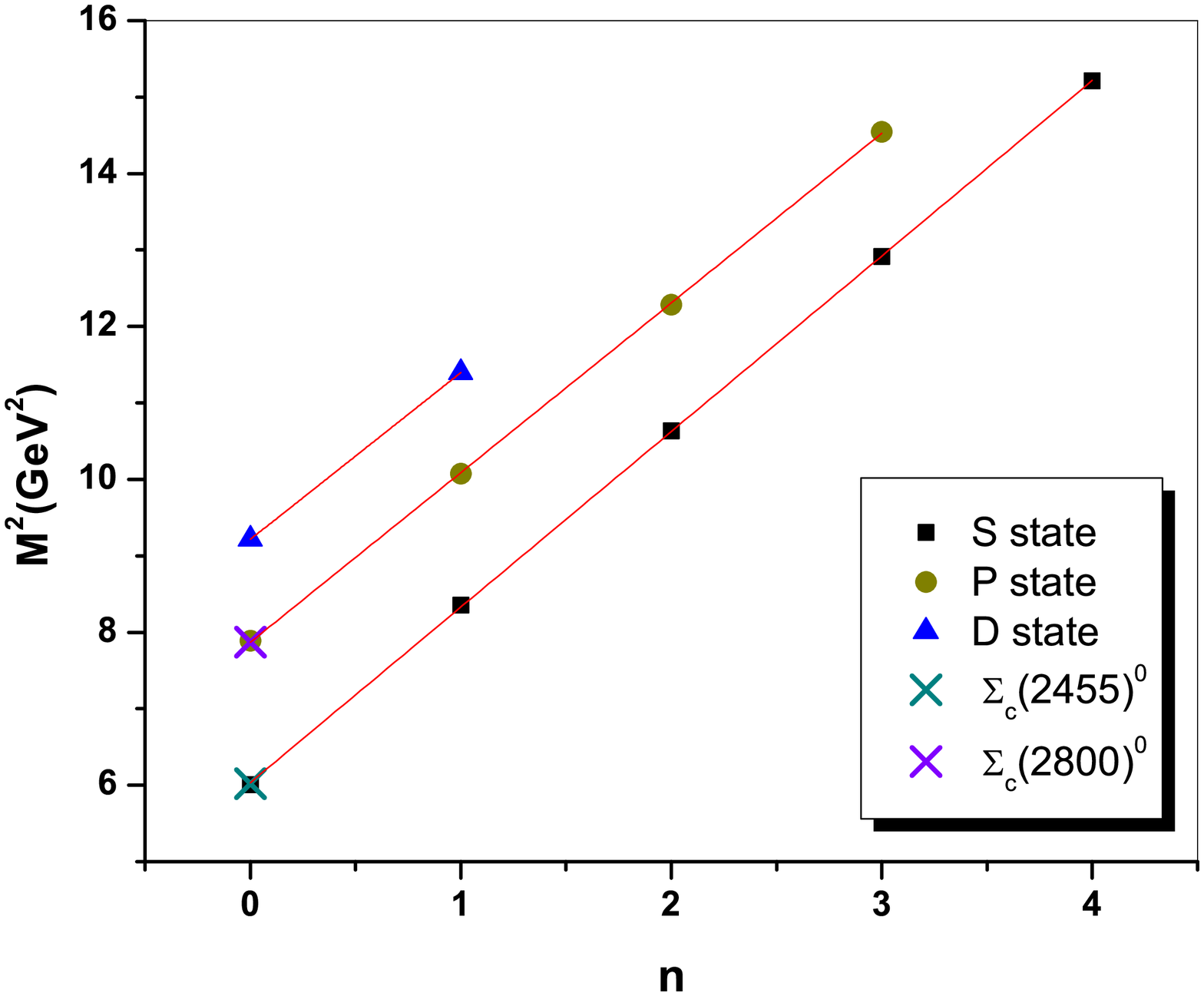}
\end{minipage}
\quad
\begin{minipage}[b]{0.40\linewidth}
\includegraphics[scale=0.30]{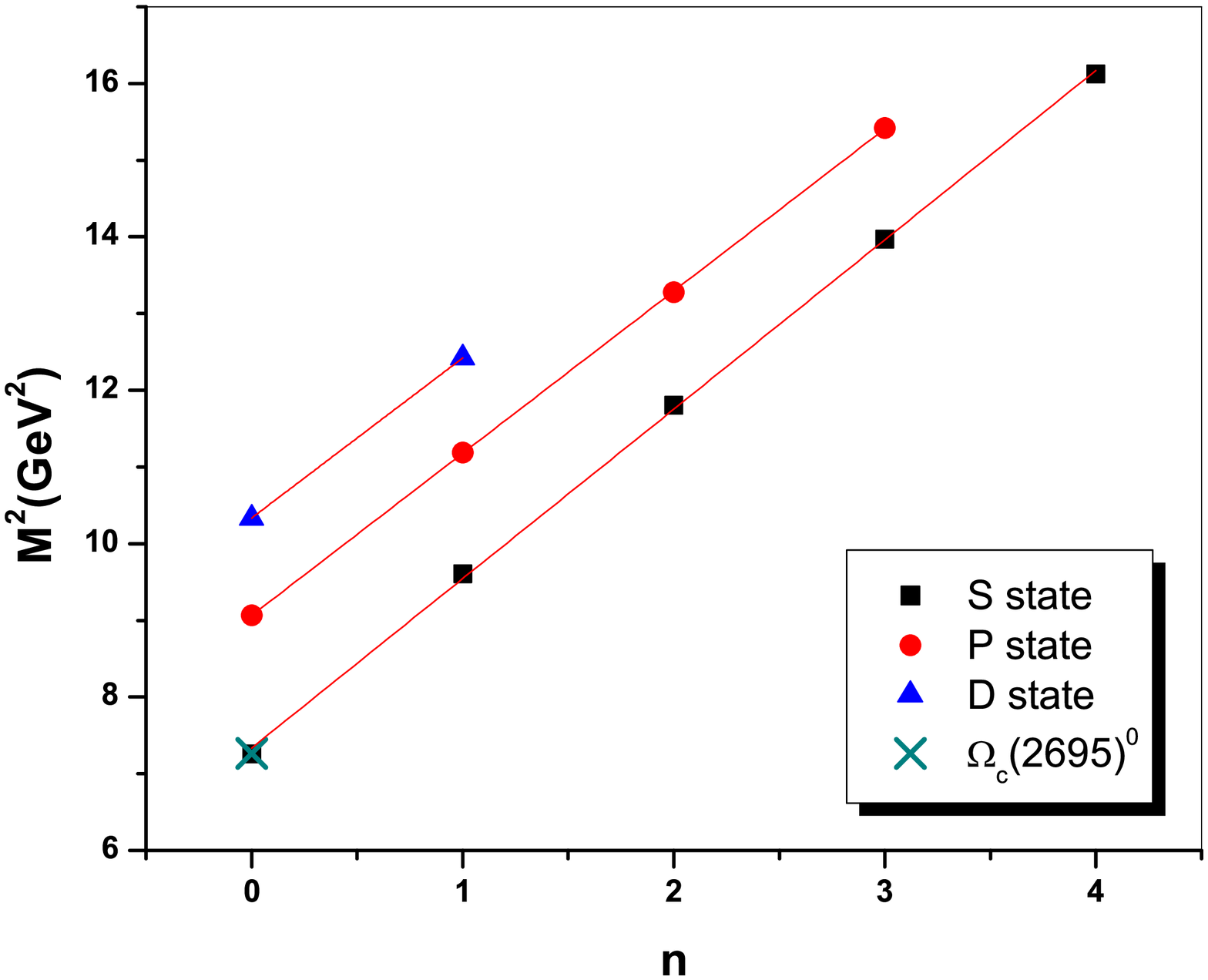}
\label{fig:2}
\end{minipage}
\caption{\label{fig:epsart} Regge Trajectory of $\Sigma_{c}^{0}$(left) and
$\Omega_{c}^{+}$(right) for $J^{P}$ values $\frac{1}{2}^{+}$,$\frac{1}{2}^{-}$ and $\frac{5}{2}^{+}$ with different states. Available Expt. states are given with particle name.}
\end{figure*}  

\section{Magnetic moments and Radiative decays}
The magnetic moment of baryons are obtained in terms of the spin, charge and effective mass of the bound quarks as \cite{bhavin, 93,95,108,15}
\begin{eqnarray}\nonumber
\mu_{B}=\sum_{i}\langle \phi_{sf}\vert \mu_{iz}\vert\phi_{sf}\rangle)
\end{eqnarray}
where
\begin{equation}
\mu_{i}=\frac{e_i \sigma_i}{2m_{i}^{eff}}
\end{equation}
$e_i$ is a charge and $\sigma_i$ is the spin of the respective constituent quark corresponds to the spin flavor wavefunction of the baryonic state.  The effective mass for each of the constituting quark $m_{i}^{eff}$ can be defined as
\begin{equation}
m_{i}^{eff}= m_i\left( 1+ \frac{\langle H \rangle}{\sum_{i} m_i} \right)
\end{equation}
where, $\langle H \rangle$ = E + $\langle V_{spin} \rangle$. 

\begin{table*}
\begin{center}
\tabcaption{\label{tab:table10} Magnetic Moment(in nuclear magnetons) of $J^{P}$~ $\frac{1}{2}^{+}$ and $\frac{3}{2}^{+}$ singly charmed baryons.}
\begin{tabular}{ccccccc}
\toprule Baryons& function&Our&\cite{103}&\cite{bhavin}&\cite{102}&\cite{101}\\
\hline
$\Lambda_c^{0}$&$ \mu_{c}$&0.422&0.411&0.385&0.39\\
$\Sigma_c^0$&$\frac{4}{3} \mu_{d}$-$\frac{1}{3} \mu_{c}$&-1.091&-1.043&-1.015&-1.60 \\
$\Sigma_b^{* 0}$& 2$\mu_{d}$+$ \mu_{c}$&-1.017&-0.958&-0.850&-1.99&-1.18 \\
  $\Xi_c^0$&$\frac{2}{3} \mu_{d}$+$\frac{2}{3} \mu_{s}$-$\frac{1}{3} \mu_{c}$&-1.011& -0.914\\
  $\Xi_c^{* 0}$& $\mu_{d}$+$\mu_{s}$+$ \mu_{b}$&-0.825& -0.746&-0.690&-1.49&-1.020\\
 $\Omega_c^-$&$\frac{4}{3} \mu_{s}$-$\frac{1}{3} \mu_{c}$&-0.842&-0.774&-0.960&-0.900\\
  $\Omega_c^{* -}$&2 $\mu_{s}$+ $\mu_{c}$&-0.625&-0.547&-0.867&-0.860&-0.840\\
  \bottomrule
\end{tabular}
\end{center}
\end{table*}

The electromagnetic radiative decay width can be expressed in terms of the radiative
transition magnetic moment(in $\mu_N$) and photon energy (k) as \cite{95,106}

\begin{equation}
\gamma_r= \frac{k^{3}}{4\pi} \frac{2}{2J+1} \frac{e^2}{m_{p}^{2}} \mu^{2}_{B \rightarrow B^{'}}
\end{equation}
Here, $m_p$ is the proton mass. $m_{B} \rightarrow m_{B^{'}}$ is the radiative transition magnetic moments (in nuclear magnetons), which are expressed in terms of the magnetic moments of the constituting quarks ($\mu_i$) of the initial and final state of the baryon. The radiative transition magnetic moment is calculated as(in terms of keV) \cite{95, bhavin}\\
\begin{itemize}
\item{ $\Sigma_c^{* 0}$ $\rightarrow$ $\Sigma_c^{0}$~~~ $\frac{2 \sqrt{2}}{3} (\mu_d-\mu_c)$~~~~-1.037}
\item{$\Xi_c^{* 0}$ $\rightarrow$ $\Xi_c^{0}$~~~ $\frac{\sqrt{2}}{\sqrt{3}} (\mu_d-\mu_s)$~~~~~-0.182}
\item{$\Omega_c^{* 0}$ $\rightarrow$ $\Omega_c^{0}$~~~ $\frac{2 \sqrt{2}}{3} (\mu_u-\mu_s)$~~~~-0.876}
\item{$\Sigma_c^{0}$ $\rightarrow$ $\Lambda_c^{+}$~~~ $\frac{\sqrt{2}}{\sqrt{3}} (\mu_u-\mu_d)$~~~~~~1.844}
\end{itemize}

\noindent Using this radiative transition magnetic moment for particular baryons, we calculated radiative decay width and the results are tabulated in Table \ref{tab:table11}.\\
\begin{center}
\tabcaption{\label{tab:table11} Radiative decay widths (in keV).}
\begin{tabular*}{80mm}{c@{\extracolsep{\fill}}cccc}
\toprule Decay&our&\cite{103}&\cite{104}&\cite{105}\\
\hline
$\Sigma_c^{* 0}$ $\rightarrow$ $\Sigma_c^{0}$&1.553&1.080&2.670&0.08 $\pm$ 0.03\\
$\Xi_c^{* 0}$ $\rightarrow$ $\Xi_c^{0}$&0.906&0.908&-&0.66 $\pm$ 0.32\\
$\Omega_c^{* 0}$ $\rightarrow$ $\Omega_c^{0}$&1.441&1.070&0.850&-\\
$\Sigma_c^{* 0}$ $\rightarrow$ $\Lambda_c^{+}$&213.3&126&176.7&130$\pm$45\\
\bottomrule
\end{tabular*}
\end{center}

%
%
%

\section{Conclusion}

The mass spectra of $\Lambda_c$, $\Sigma_c$, $\Xi_c$ and $\Omega_c$ baryons are calculated in hypercentral constituent quark model(hCQM) with coulomb plus linear potential. The mass spectra of $\Lambda_c$, $\Sigma_c$, $\Xi_c$ and $\Omega_c$ baryons are close to known experimental observations and other theoretical predictions as shown in Table \ref{tab:table2}- \ref{tab:table5}. We have listed various experimentally known singly charmed baryonic states in Table~\ref{tab:table12} and compared them with our calculated results. We have also assigned $J^P$ value to unknown experimental states. We can say that there are not much theoretical calculations which provide the spectra starting from S to F states. So, we have compare all our results to the D. Ebert calculations \cite{ebert2011} and they are matching upto certain energy scale. This study will help experimentalists as well as the theoreticians to understand the dynamics of the singly charmed baryons. \\

\begin{center}
\tabcaption{\label{tab:table12} The comparison between our predicted baryonic states and experimental unknown($J^P$) excited states.}
\begin{tabular*}{80mm}{c@{\extracolsep{\fill}}cc}
\toprule Names & M(GeV)\cite{olive}& Baryon State \\
\hline
$\Lambda_{c}^{+}(2765)$ &2766.6$\pm$2.4&$(2^2S_\frac{1}{2})$\\
$\Lambda_{c}^{+}(2940)$&2939.3$^{+1.4}_{-1.5}$& $(1^2P_\frac{1}{2})$,$(1^2P_\frac{3}{2})$ \\
$\Sigma_{c}^{0}(2800)$ & $2806 \pm8\pm10$&$(1^2P_\frac{1}{2})$,$(1^2P_\frac{3}{2})$,$(1^4P_\frac{5}{2})$\\
$\Xi_{c}(2930)$ & 2931$\pm$3$\pm$5 &$(1^2P_\frac{1}{2})$,$(1^2P_\frac{3}{2})$\\
$\Xi_{c}^{0}(2980)$ & 2968.0$\pm$2.6&$(2^2S_\frac{1}{2})$,$(2^4S_\frac{3}{2})$\\
$\Xi_{c}^{0}(3080)$ & 3079.9$\pm$1.4&$(1^2D_\frac{3}{2})$, $(1^4D_\frac{3}{2})$ \\
\bottomrule
\end{tabular*}
\end{center}

\par The higher excited orbital and radial states mass calculation allow us to construct the Regge trajectory in ($n_{r}$, $M^{2}$) plane. The experimental masses in Fig. (1-2) are near to the obtained masses such that, we are getting almost linear, parallel and equidistance lines in each case of baryons. In Fig.[1] 1S and 1D states of $\Lambda_{c}^{+}$ are matched with our results while our 1P state is higher than experimental value.
$\Lambda_{c}(2765)^{+}$ is perfectly matched with our 2S state and $\Lambda_{c}^{+}(2940)$ can be assigned as 2P. On other side, 1S, 1P and 1D states of $\Xi_{c}^{0}$ are reasonably close to obtained results and $\Xi_{c}(2990)^{0}$ value is nearer to 2S state, thus we assigned it as 2S state. In Fig. [2] $\Sigma_{c}(2455)^{0}$ and $\Sigma_{c}(2800)^{0}$ are exactly coincide with our 1S and 1P states. Our results of $\Sigma_c^{0}$ are pretty much close to theoretical models and experiments in comparisons to other obtained baryon states. The only (experimental available) ground state value of $\Omega_{c}^{0}$ is matched with our 1S state. \\

The magnetic moments and radiative decay widhts are also cxalculated for ground state baryons with $J^{P}$~ $\frac{1}{2}^{+}$ and $\frac{3}{2}^{+}$. The obtained results are close to the other theoretical prediction. Thus, our motive to calculate the excited states has been fulfilled. We can observe that the obtained masses of four singly charmed baryons have also been checked by bound states baryonic properties; regge trajectories and radiative decay widths. The model is succeed to determine the properties, thus, we would like to extend this scheme to calculate the decay rates of these baryons. We will also implant the model in calculation of doubly and triply heavy baryons properties.\\

%
%
%
%
%

\vspace{1mm}

\acknowledgments{One of the author Z. Shah would like to thank Dr. Nilmani Mathur for his valuable suggestions at CCP 2015 at IIT, Guwahati. A. K. Rai acknowledges the financial support extended by DST, India  under SERB fast track scheme SR/FTP /PS-152/2012. }

\end{multicols}

\vspace{-1mm}
\centerline{\rule{80mm}{0.1pt}}
\vspace{2mm}

\begin{multicols}{2}

\end{multicols}

\clearpage

\end{document}